\documentclass[%
 reprint,
 amsmath,amssymb,
 aps,
showkeys,
]{revtex4-2}

\usepackage{graphicx}
\usepackage{dcolumn}
\usepackage{bm}
\usepackage{hyperref}
\usepackage[hyphenbreaks]{breakurl}

\usepackage{orcidlink}

\usepackage{tikz}
\usetikzlibrary{shapes.geometric, arrows}
\tikzstyle{stage} = [rectangle, rounded corners, 
minimum width=1cm,
minimum height=1cm,
text centered, 
text width=0.25\textwidth,
draw=black]
\tikzstyle{stage1} = [rectangle, rounded corners, 
minimum width=1cm,
minimum height=1cm,
text centered, 
text width=0.8\textwidth,
draw=black]
\tikzstyle{result} = [diamond, 
minimum width=1cm, 
minimum height=1cm, 
text centered, 
draw=black, 
fill=green!30]
\tikzstyle{arrow} = [ultra thick,->,>=stealth]

\begin{document}

\preprint{APS/123-QED}

\title{Global transformer overheating from geomagnetic storms}

\author{Morgan Rivers \orcidlink{0000-0001-9367-8280}}
 \affiliation{Alliance to Feed the Earth in Disasters (ALLFED), 603 S. Public Rd \#57 Lafayette, CO 80026, USA}
 \email{morgan@allfed.info}
\author{Łukasz G. Gajewski \orcidlink{0000-0003-3097-0131}}%
\affiliation{Alliance to Feed the Earth in Disasters (ALLFED), 603 S. Public Rd \#57 Lafayette, CO 80026, USA}%
\author{David Denkenberger \orcidlink{0000-0002-6773-6405}}
\affiliation{Alliance to Feed the Earth in Disasters (ALLFED), 603 S. Public Rd \#57 Lafayette, CO 80026, USA}
\affiliation{Department of Mechanical Engineering, University of Canterbury, Christchurch, Canterbury 8041, NZ}


\begin{abstract}
Geomagnetic storms occurring due to sustained, high-speed solar winds are known to induce currents in power distribution networks. These geomagnetically induced currents (GICs) can cause high voltage transformers (HVT) to overheat, thus resulting in a catastrophic electricity loss (CEL). 
Since significant portions of infrastructures around the world rely heavily on access to electric power, it is essential to estimate the risks associated with GICs on a global scale.
We assemble multiple methodologies across various scientific disciplines to develop a framework assessing the probability of a severe geomagnetic storm causing a long-term, widespread power outage.
Our model incorporates thermal models of HVT tie bar hot spots, historical geoelectric field estimates, and a global conductivity model to estimate the risk of long-term power outage for regions between $-70^\circ$ and $80^\circ$ geomagnetic latitude due to transformer overheating failure.
Assuming a uniform $33\%$ HVT spare capacity, our analysis indicates that a 1 in 10,000 year storm would result in approximately $1\%$ of the population in Europe and North America experiencing a long-term (months to years) electricity loss.
\end{abstract}

\maketitle


\section{Introduction}
We have submitted this work to a peer-reviewed journal and received insightful feedback on our modelling approach. The assigned reviewers pointed out certain flaws and limitations in our model that ultimately prevented this paper from being published without a major revision.

As an organisation with a constrained budget, we are unfortunately forced to prioritise and focus on our mission to help build resilience to global catastrophic food system failure. Therefore, we cannot afford the further investment to address the revisions.
While the corrections recommended by the reviewers would alter the results (especially regionally), we do not believe that this change would imply that a geomagnetic storm is a likely cause of global catastrophic food failure.
However, we do encourage the scientific community to utilise this work, our open-source code repository, and the reviewers' comments to establish a better version of what we endeavoured to achieve here.

We paraphrase the received feedback, focusing on the most critical issues:

There is a significant concern with the quality of data we used as the foundation for the electromagnetic fields and the use of the Alekseev et al., model \cite{alekseev2015compilation}, which both result in an overly smooth picture of the electromagnetic fields. Thus, our predictions for specific regions could be incorrect by orders of magnitude due to the lack of information about local variability in our approach.
While there are now newer datasets that we could utilise, they are only available for the US and Australia; therefore, to provide a global scale prediction, one would need to extrapolate from those, which could create its own problems.
An additional concern is the lack of long-term geomagnetic data available to predict extreme statistics for time spans as long as 1-in-10000 years. 
While this is true, the reason we included such a large time window is that a) it is largely absent in the contemporary literature on this topic, and b) in near-term predictions, there is little to no risk of a global catastrophe, which is our priority concern.
Little risk of a global catastrophe does not mean, however, that there is no substantial regional risk due to geomagnetic storms; thus, the aforementioned encouragement for other researchers to pick up this mantle.
We would also like to point out that a small but significant loss in electricity could lead to a much greater loss in both regional and global industrial output \cite{blouin}.
Moreover, we made certain oversimplifications that should eventually be addressed, such as assuming the dominant East-West direction of EM fields, uniform ground resistances, GIC per phase values and similar power grid characteristics worldwide.
Finally, and of equal importance,  our method is, in essence, a statistical analysis built upon a chain of approximations, and rigorous uncertainty analysis is a challenge yet to be met.

Additional references suggested by the reviewers: \cite{bedrosian2015mapping, kelbert2020role, schulte2014severe, eastwood2018quantifying}.

The rest of the manuscript that follows is in its unaltered state from what we submitted for review.

The phenomenon known as the geomagnetic storms has been studied for hundreds of years\cite{lakhina2016geomagnetic}, but only recently, due to advances in electrification, some are starting to worry about the potential risks they can pose to our society \cite{organek2017black, shulz}.
A society that is increasingly dependent on electricity and related technologies \cite{organek2017black, lofquist2020there}.
\begin{figure*}[t]
\centering
\resizebox{1\textwidth}{!}{
\begin{tikzpicture}[node distance=2cm]

\node (stage1) [stage1] {Geoelectric Fields for Severe Storm\\
\begin{itemize}
    \item Historical $B$ field levels.
    \item Apparent conductivity and resistivity at TF sites.
    \item $E$ field and rates of storm recurrence.
    \item Log-normal fit to adjusted $E$ field levels to an equivalent field at a reference location.
    \item Global $E$ field levels.
\end{itemize}
};

\node (stage2) [stage, below of=stage1, yshift=-2.5cm, xshift=-4.5cm] {GIC and Transformer Overheating in the High Voltage Power Distribution Network\\
\begin{itemize}
    \item GICs and transformer properties.
    \item Probability of transformer failure due to overheating.
\end{itemize}
};

\node (stage3) [stage, right of=stage2, xshift=3.5cm] {Effects of Transformer Failure on the Population and Electricity Grid\\
\begin{itemize}
    \item Regions supplied by transformer nodes.
    \item Electricity Loss for each region.
\end{itemize}
};

\node (final) [result, right of=stage3, xshift=2.5cm] {Final results.};

\draw [arrow] (stage1) -- (stage2);
\draw [arrow] (stage2) -- (stage3);
\draw [arrow] (stage3) -- (final);

\end{tikzpicture}
}
\caption{Simplified flowchart of our proposed framework. The general model consists of three main stages and results in concrete estimates of regional power outages for arbitrary locations on Earth and how those outages affect the population. A detailed flowchart is available along with the model's source code at \href{https://github.com/allfed/GeomagneticModel}{https://github.com/allfed/GeomagneticModel}.}
\label{fig:flowchart}
\end{figure*}
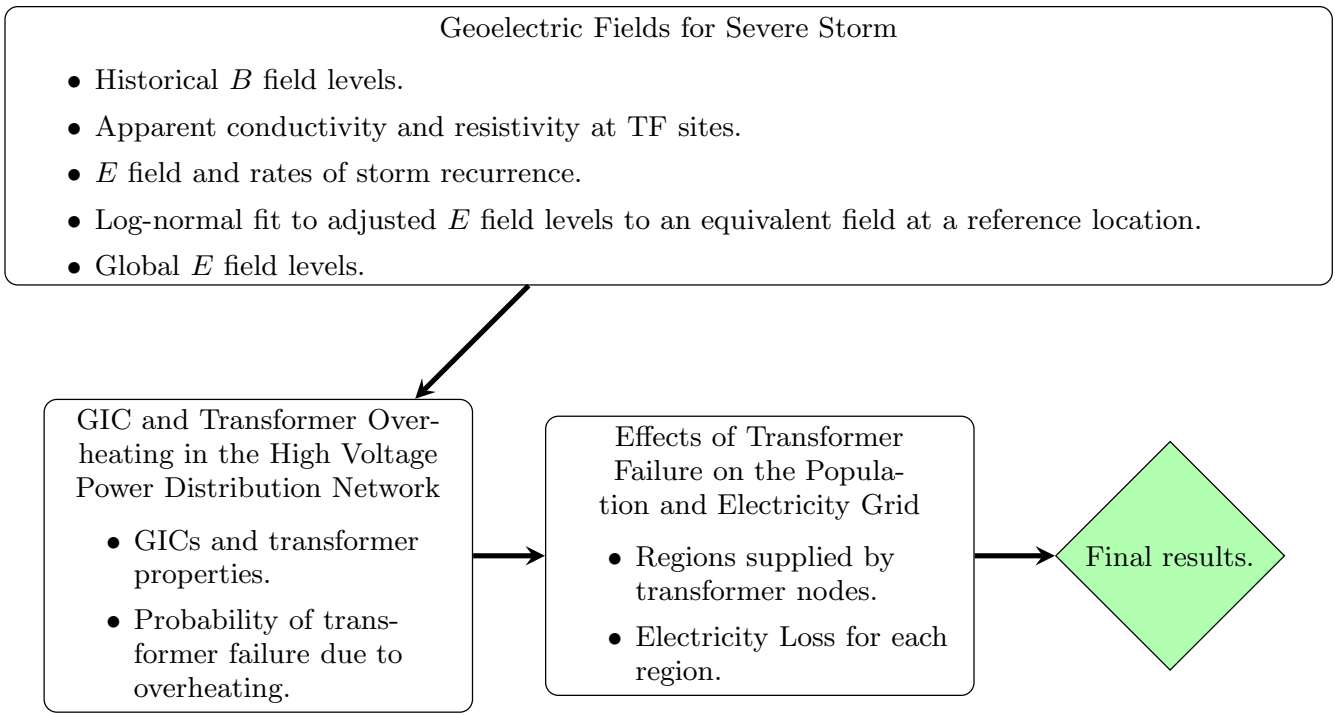
Geomagnetic storms result from an energy exchange from the solar wind into Earth's magnetosphere. They are triggered by sustained periods of high-speed solar wind, and the most significant storms involve solar coronal mass ejections (CMEs) \cite{gosling1993solar}.
CMEs are the largest-scale eruptive phenomenon in the solar system -- the bulk of plasma with a mass of $10^{13}$ kg is hauled up out to the interplanetary space with a velocity of more than 1000 km/s \cite{chen2011coronal}.
These storms induce intense currents in Earth's magnetosphere, affecting the radiation belts, ionosphere, and thermosphere. We typically measure them by indices such as the disturbance storm time -- Dst -- and a planetary geomagnetic disturbance index -- Kp \cite{gonzalez1994geomagnetic}. 
The consequences include increased atmospheric density affecting satellite orbits, disruptions to navigation systems like GPS, and the generation of harmful geomagnetic induced currents (GICs) in power grids. 
These extreme events can also be visually stunning in the form of aurorae, such as the commonly known northern lights, despite their potentially disruptive nature \cite{noaa-geomag}.

GICs are what they sound like -- electric currents that appear (are induced) in, e.g., power lines as a consequence of rapid changes in Earth's magnetic field, and those currents can be a cause of worry for the stability and health of power distribution networks \cite{pulkkinen2017geomagnetically}.

This worry about GICs mainly consists of two phenomena: overheating components (predominantly transformers) and tripping safety switches causing cascade failures \cite{roodman2015risk}.
The jury on whether these constitute a significant societal risk is still out.
Scientific studies on this issue are mostly localised (i.e., consider one particular region instead of all of the Earth), and severity claims vary from no risk at all \cite{vsvanda2021modelling, girgis2012effects} to catastrophic, such as trillions of dollars in damages and years of recovery \cite{national2008severe, chrissaid, weiss2019assessment}, with most ending somewhere in between \cite{torta2014assessing, gil2021evaluating, risa.13229, roodman2015risk, gaunt2007transformer, moodley2012developing}.
It is worth noting that while the global risk of GICs is still undetermined, there are documented cases of damage to power grids from GICs \cite{gaunt2007transformer, moodley2012developing}, including severe cases like the Quebec power failure of $1989$ \cite{gaunt2016space}. 
The problem with assessing the risk here is that strong enough events to cause damage are exceedingly rare, and a replay of the largest documented case from $1859$, commonly referred to as the Carrington event\cite{tsurutani2003extreme}, is yet to be seen.

For further reading on this topic, we refer the Reader to \cite{risa.13229, roodman2015risk, gaunt2016space} and references therein.

Since there seems to be a gap in the body of literature dealing with widespread events reaching global scales, we attempt to partially fill this gap in this paper.
Even though our analysis here shall be limited to Europe and North America due to limits in data availability, we develop a methodology to assess the consequences of various strengths of geomagnetic storms on a global scale.
The framework can be applied to any geographical region where the power grid is known with sufficient accuracy and completeness.
The software implementation of our model is open-source\footnote{\href{https://github.com/allfed/GeomagneticModel}{https://github.com/allfed/GeomagneticModel}}, written in Python, and can be run on a modern PC without the need for a computing cluster or a super-computer.

In this article, we specifically tackle the issue of overheating transformers in power grids due to GICs on an international scale, utilising data from global magnetic field observation (MT) stations \cite{gjerloev2009global}, electromagnetic transfer function (EMTF) repository \cite{kelbert2011iris,kelbert2019taking, kelbert2019first}, crowd-sourced power distribution network \cite{rivera2017opengridmap} as well as OpenStreetMap \cite{OpenStreetMap}, North American Electric Reliability Corporation (NERC) \cite{nercEOP-010-1, nerc2013}, and Electric Power Research Institute (EPRI) \cite{epri2020, epri2017, epri2019, Kappenman2010MetaR322LP}.
In order to assess the effect on population, we used data from the U.S. Energy Information Administration, and the Center for International Earth Science Information Network, Columbia University \cite{popcountry, popdata, electrcountry}.

Our primary target is estimating the population numbers experiencing a catastrophic electricity loss (CEL) given a recurrence class of a geomagnetic storm (e.g., a $1$ in $100$ year event).

\section{Methods}

In Fig.~\ref{fig:flowchart}, we present a simplified flowchart of our model (a full picture is available in the source code repository: \href{https://github.com/allfed/GeomagneticModel}{https://github.com/allfed/GeomagneticModel}).
We partition the model into three distinct stages, described in detail as follows.

\subsection{Stage I} 
This stage is based on the methodology developed by Love et al., \cite{love2018geoelectric} and consists of estimating historical magnetic field -- $B$ -- levels, determining apparent conductivity and resistivity at electromagnetic transfer function sites (EMTF or TF, for short), calculating electric field -- $E$ -- levels at magnetotelluric (MT) stations and recurrence rates for those levels, adjusting $E$ field levels to a reference location, and finally computing $E$ fields around the globe from the adjusted $E$ fields.

MT stations have been measuring the $B$ field regularly for many decades. Through these data, we are able to determine the statistics on the recurrence of $E$ fields over a long period and thus forecast the repeat rates of the less frequent, larger amplitude events. 
We obtained the data at the Fresno, California, USA MT site (FRN)
from the global ground-based magnetometer initiative (SuperMAG) \cite{gjerloev2009global}.
These data contain measurements of the magnetic field every minute for $37$ years ending in $2019$. 
It is worth noting that the results presented here depend on the choice of the site due to inaccuracies in the apparent conductivity model.
The FRN dataset was picked because of its proximity to an EMTF site (RET06), sufficient data size, and time span.
For more precise results, data from all existing MT and EMTF sites would need to be combined into one dataset or a better apparent conductivity model established, ideally both; however, this falls out of the scope of this paper.

We then process this dataset as described in \cite{love2018geoelectric} Sec. $5$: the time series of $B$ is detrended by subtracting a second-order polynomial fitted to the entire duration of the time series, and any missing data are filled by linear interpolation. Then, the series is put through a fast Fourier transform (FFT), multiplied by an appropriate transfer tensor for the survey site, and finally, a reverse FFT is performed, resulting in a geoelectric -- $E$ -- series.
This procedure is rooted in the relationship between the geomagnetic and geoelectric fields at the Earth's surface: \cite{berdichevsky2008models, chave2012magnetotelluric}
\begin{equation}
    \mathbf{E}(f, x, y) = \frac{1}{\mu}\mathbf{Z}(f, x, y) \cdot \mathbf{B}(f, x, y),
    \label{eq:geomagele_relationship}
\end{equation}
where $f$ is the frequency of sinusoidal variation, $x$ and $y$ are horizontal Cartesian components, $\mu$ is permeability, and $\mathbf{Z}$ is the impedance tensor specifying the amplitude, polarisation and phase of the geoelectric field. The transfer tensor mentioned earlier is $\mathbf{Z}/\mu$.

In order to compute $E$ fields using Eq.\eqref{eq:geomagele_relationship} we need to determine the apparent conductivity and resistivity at the corresponding TF site. 
We acquire the necessary data from the SPUD EMTF repository \cite{kelbert2011iris,kelbert2019taking, kelbert2019first} at the Dog Creek, California, USA site (RET06) corresponding to the chosen MT site in Fresno.
The apparent conductivity can be expressed as: \cite{love2018geoelectric}
\begin{equation}
    \sigma_A (f) = \frac{2\pi\mu f}{|\mathbf{Z}(f)|^2},
    \label{eq:conduct}
\end{equation}
where $|\cdot|$ is the Frobenius norm\cite{berdichevsky2008models}, and the apparent resistivity is given by: \cite{love2019extreme, berdichevsky2008models}
\begin{equation}
    \boldsymbol{\rho} (f) = \frac{|\mathbf{Z} (f)|_\odot^2}{2\pi\mu f},
    \label{eq:resist}
\end{equation}
where $|\cdot|_\odot$ represents an element-wise norm.

With $E$ fields established, we now consider an adjustment coefficient that describes the relationship between fields' strengths and durations (SI Fig. S1).
The peak electric field levels that occur once every ten years change with the duration over which they are sustained; namely, longer duration corresponds to lower peak field levels. 
Thus, we compare the mean field level lasting sixty seconds to longer-lasting fields.
We shall use this adjustment later to predict the strength of GICs in transformers to ensure we do not overestimate the current levels of long-duration GICs.

Our ultimate goal of this stage is to compute $E$ fields (their strength and occurrence rates) at an arbitrary location on Earth, extrapolating from the measurement site. 
For this purpose, we adjust all measured fields to an arbitrarily chosen reference with conductivity $\sigma_{ref} = 8.7*10^{-6} [\frac{\text{S}}{\text{m}}]$, located at a magnetic latitude $L_M = 72^\circ$ N so that we can model $E$ fields globally with a single function regardless of the choice of the MT site or even ensemble of sites.
To this end, we incorporate an auroral boundary shift using an estimated auroral boundary movement of $0.004$ degrees threshold shift south per [nT] Dst as per the recommendation in the report from EPRI \cite{epri2020}, with the Dst estimates taken from Oughton et al., \cite{risa.13229}.
Subsequently, we use the geomagnetic latitude distribution data from Ngwira et al.,\cite{ngwira2013extended} to compute a magnetic latitude adjustment $\hat{L_M}$, and we establish the apparent conductivity adjustment as $\hat{\sigma_A} = \sigma_A / \sigma_A^{ref}$.

Thus, we obtain a formula for the reference $E$ field:
\begin{equation}
    |E_{ref}| = \hat{L_M} \hat{\sigma_A} |E|.
    \label{eq:eref}
\end{equation}
We can now fit a log-normal function to all $E$ field data (SI Fig. S2) and then use this model to estimate geoelectric fields at an arbitrary point on Earth \cite{love2020some, love2018geoelectric}.

This arbitrary location $E$ field estimation is conducted with the help of Earth's ground conductivity at a $15$ arc-minute basis from Alekseev et al., \cite{alekseev2015compilation}. It is a layered model, so we calculate the impedance recursively as prescribed in the NERC report \cite{nerc2013}.
Finally, we adjust the E field at the reference site back to the expected field levels for all points on Earth at $15$ arc-minute resolution via:
\begin{equation}
    |E|=\frac{1}{\hat{L_M}\hat{\sigma_A}}|E_{ref}|.
    \label{eq:efield}
\end{equation}

\begin{figure}
    \centering
    \includegraphics[width=\linewidth]{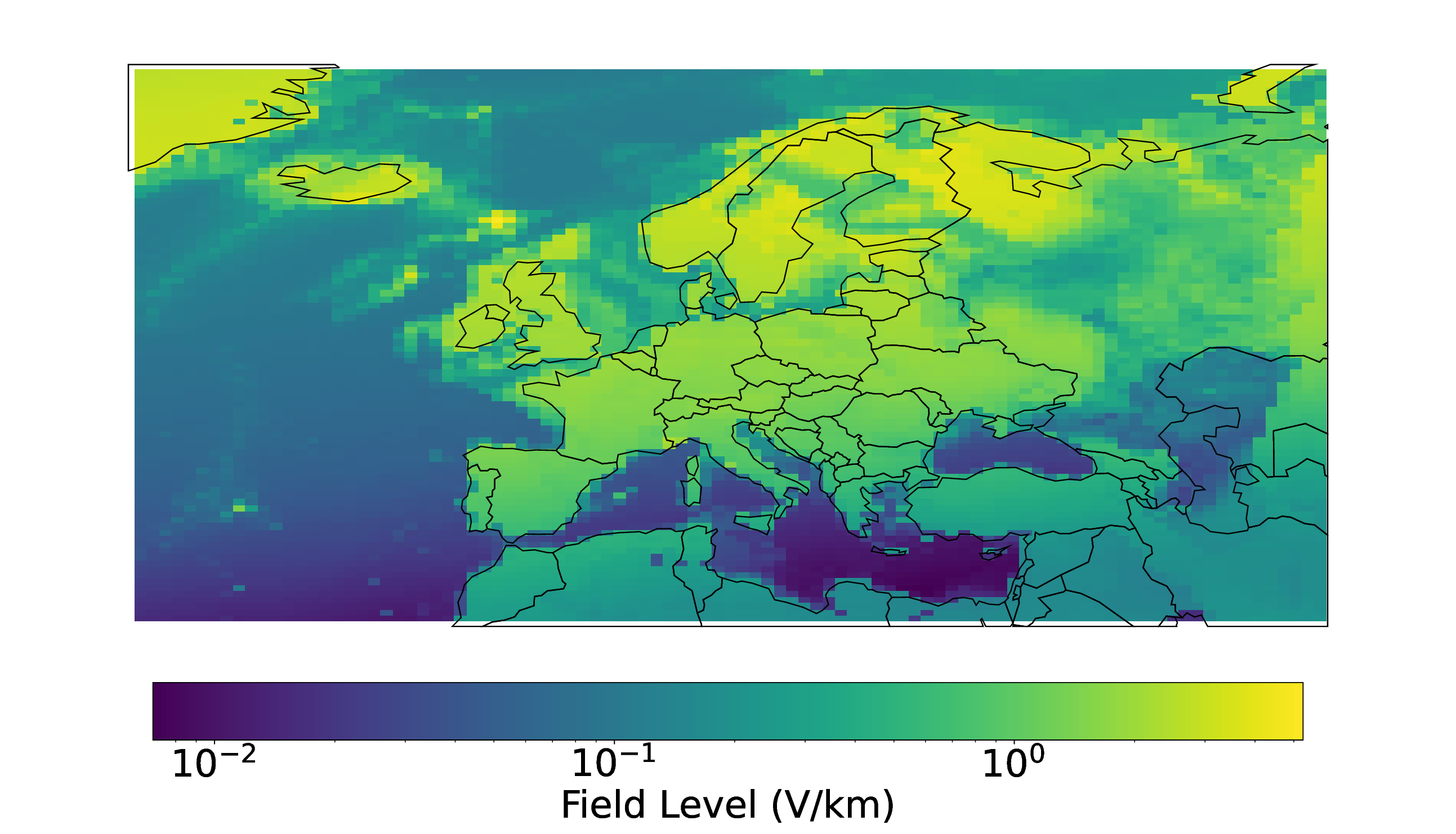}
    \includegraphics[width=\linewidth]{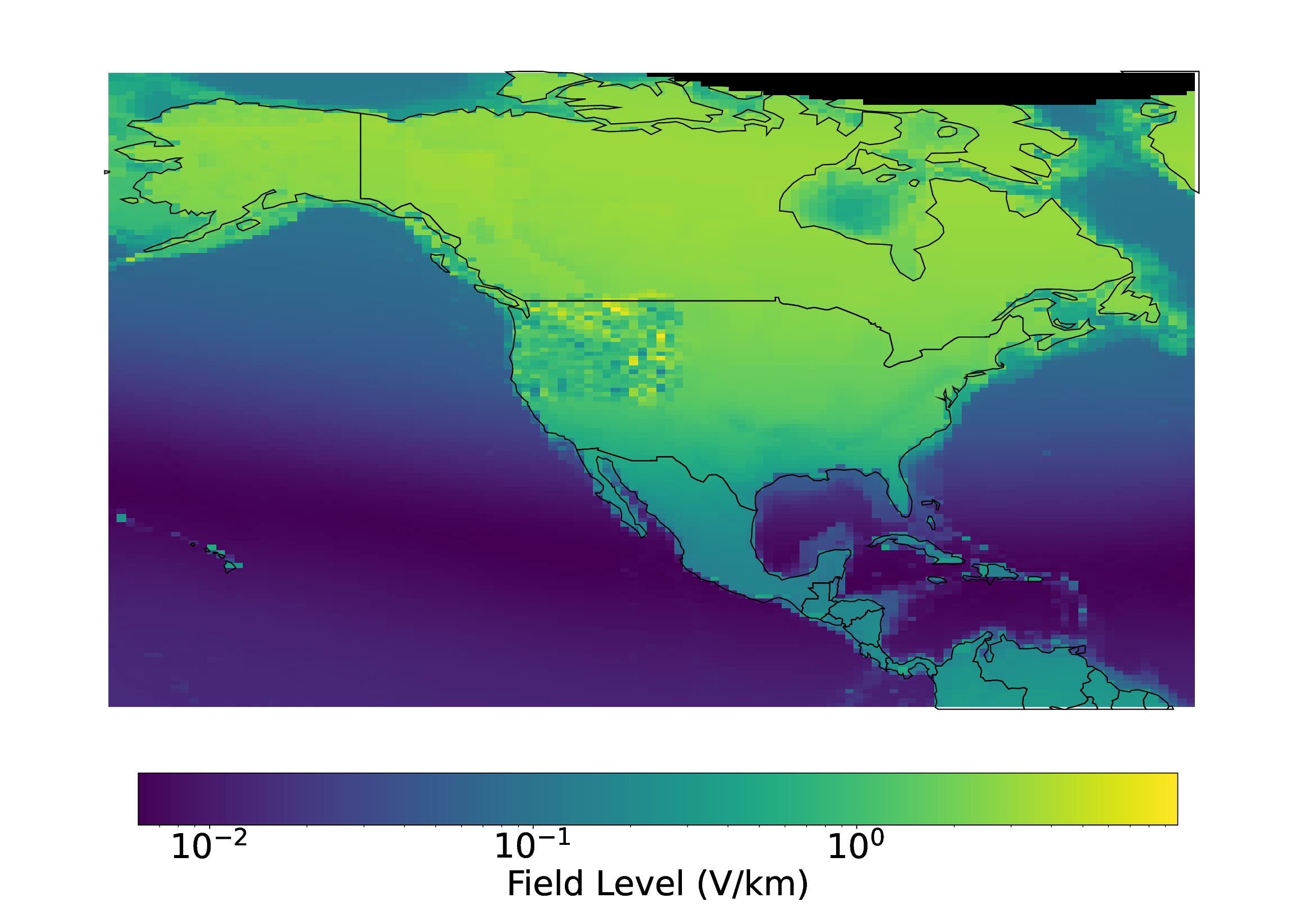}
    \caption{Peak $60$-second geoelectric field magnitude for a $1$-in-$100$ year storm in Europe (top) and North America (bottom). The colour indicates the field level [V/km].}
    \label{fig:efields}
\end{figure}

This concludes stage one, and in Fig.~\ref{fig:efields}, we present our prediction for the expected peak magnitude of a sixty-second $E$ field in a $1$-in-$100$ year storm for Europe and North America.

\begin{figure*}[t]
    \centering
    \includegraphics[width=1\linewidth]{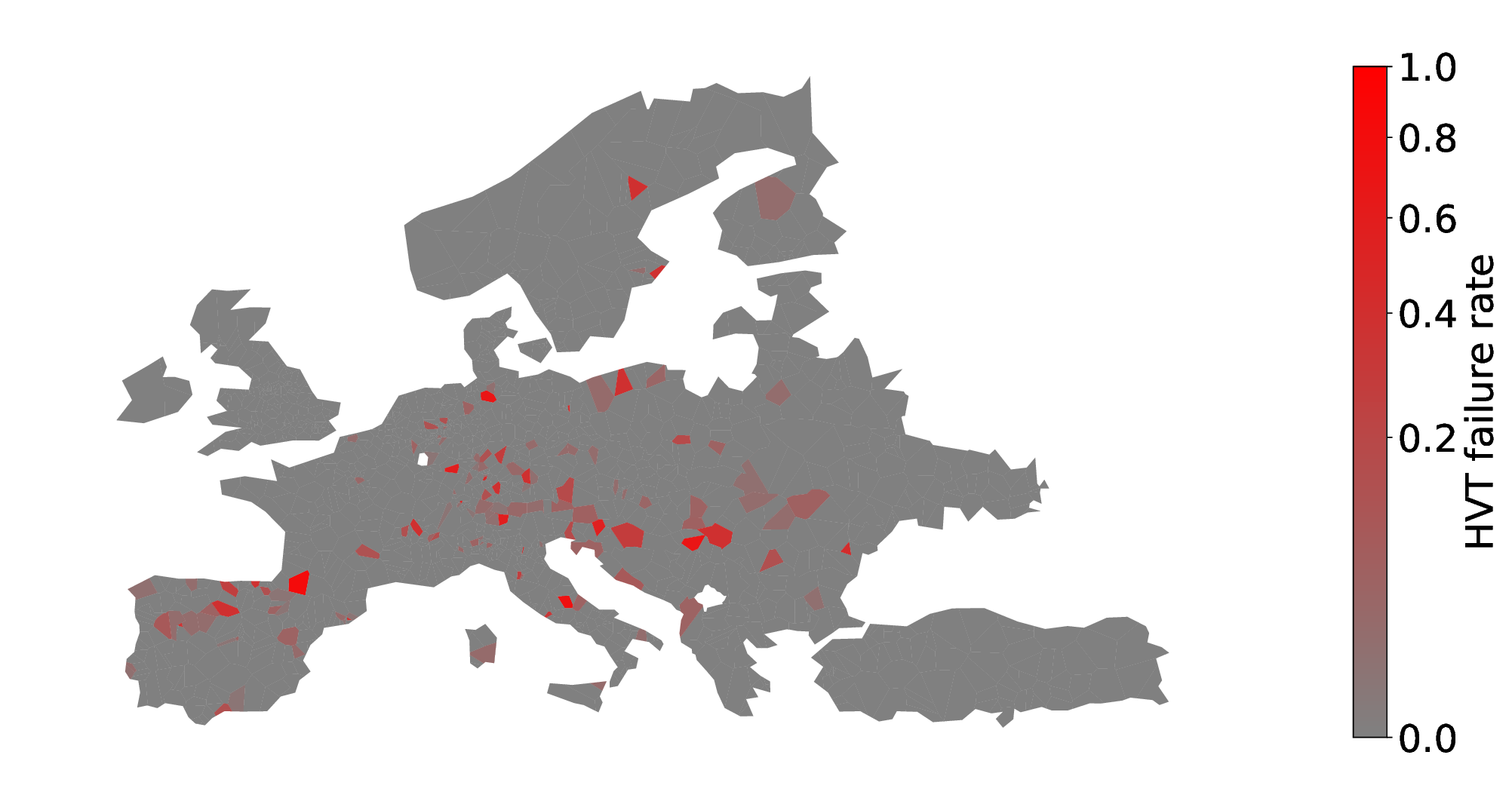}
    \caption{European substations at risk of electricity loss after a $1$-in-$10000$ year storm. Colour indicates the fraction of HVTs we expect to fail within a given Voronoi tile.}
    \label{fig:euvoronoi}
\end{figure*}

\subsection{Stage II} 
In this stage, we determine the GICs in power grids and, thus, the transformer failure due to overheating.
We gathered the worldwide power distribution network data from OpenGridMap \cite{rivera2017opengridmap}.
It is an open-source initiative that utilises crowd-sourced data to automatically generate power grid models in a Common Information Model (CIM) format, which is the power industry standard \cite{simmins2011impact}.
Since we focus in this paper on high voltage networks and high voltage transformers (HVT), we selected all lines above $100$ [kV].
As pointed out in the introduction, we will be restricting ourselves to only two continents, and the primary reason for this is data availability (or lack thereof) in the OpenGridMap.
We present plots of the selected networks in SI Fig. S3.

On these grids, the GICs were computed with the help of the open-source package -- GEOMAGICA \cite{geomagica}, which implements the Lehtinen-Pirjola method \cite{Lehtinen1985CurrentsPI} for the Austrian grid. 
We modified this code such that it can be applied to an arbitrary location on Earth.
Each power line length was preserved accurately, and the full length of the line was used to estimate overall line resistance; however, the GEOMAGICA package only allows for straight line estimates of induced GIC. 
The resistance per kilometre, as well as HVT winding conductance, were estimated using linear interpolation of line conductance as a function of AC voltage using values provided in a NERC report \cite{nercEOP-010-1}, with the ground resistance being set to $0.5 [\Omega]$.

Final GIC per phase values for transformers were estimated as the expected GIC per phase flowing through the ground grid divided by $1.3$. This factor brings the final values closer to those obtained with more complex transformer circuit configurations and is discussed in further detail in \cite{risa.13229}.
The absolute values of geoelectric fields were applied uniformly in the East-West direction, as this is the primary field orientation \cite{kikuchi2008penetration, marshall2013observations}.
If the GIC duration is longer than a minute, we scale its magnitude down to account for the decreasing statistical likelihood of an $E$ field sustained for a longer duration (as explained in stage one, also see, SI Fig. S1), and the fact that GICs are proportional to the electric field $I_{GI} \propto E$.

With GIC levels and their recurrence rates established, we move on to determine whether they cause sufficient heat increase in transformers to result in transformer failure.
We source the steady state temperature rise $T^*(I_{GI})$ for $42$ different types of transformers and two tie bar geometries from an EPRI report \cite{epri2019} and linearly interpolate the data where necessary, as well as the temperature response for both designs -- $\tau$. 
These data were extracted by hand from the said report.
Both $T^*$ and $\tau$ are multiplied by $\frac{2}{3}$ to accommodate the fact that the studied tie bar geometry designs represent the extreme ``best'' and ``worst'' cases; thus, we are considering the midpoint between them (this scaling factor can be found via inspecting the EPRI report).
The HVT temperature can be found as: \cite{epri2017}
\begin{equation}
    T_{HVT} = T^*(I_{GI})\left(1 - e^{-\frac{t}{\tau}} \right) + T_{oil},
    \label{eq:hvttemp}
\end{equation}
where $t$ is the duration, and $T_{oil}=90^\circ C$ is the surface oil temperature assumed in a severe geomagnetic storm \cite{epri2019}.

To determine whether the temperature in Eq.~\eqref{eq:hvttemp} makes a transformer fail, we look at the structural temperature limit in \cite{epri2017}, which is a function of the transformer's age.
We estimate the transformer population age distribution based on the statistics in the same report.
Each transformer has an associated voltage class ($V_c$), which we assign based on the proximity to one of $\{230, 345, 500, 765\}$ [kV].
These values are chosen specifically as these have associated statistics on the transformer phase demographics in the aforementioned report and prevalence in the US grid per class \cite{Kappenman2010MetaR322LP}.
The US dataset was the easiest to obtain, and we assume grids around the world would not have significantly different characteristics in these particular features.
However, should that not be the case, it is fairly straightforward to substitute this number for a region-specific one if needed (for those who have access to the data).
We use the cumulative fraction of transformer population exceeding safe temperatures as the proxy for the probability of failure for the transformer in a given class; therefore, we can write:
\begin{equation}
    F(V_c) = f(V_c) \sum_A f(T_{HVT} > T_{max}) f(A),
\end{equation}
where $F(V_c)$ is the fraction of transformers overheating in class $V_c$, $A$ is the age category, with $f(A)$ the fraction of transformers in given age category, and $f(T_{HVT} > T_{max})$ is the fraction of transformers above safe temperature for the age, and $f(V_c)$ is the fraction associated with the transformer type.
For each of the $42$ transformer types, we associate $V_c$ and a phase $\phi$ with them, and now, each transformer is a member of a group. E.g., there are seven types in a group defined by $\{V_c=500, \phi=1\}$, so for each transformer in this group, $f(V_c)=1/7$.
All the values for transformer types, temperature limits, and transformer population are presented in SI Table S1, S2, and S3.

Ultimately, we want to assess whether a power network node $n$ fails, and we achieve that by estimating the percentage of transformers failing at any given node as:
\begin{equation}
  F(n) = \sum_{\{\phi, V_c\}} \left( f(\phi) f_{grid}(V_c) \sum_{V_c} F(V_c) \right).
  \label{eq:failure}
\end{equation}
$f_{grid}(V_c)$ is the prevalence of transformers in class $V_c$ in the power grid, and $f(\phi)$ is the fraction of transformers with phase $\phi$. 
If $F(n) > 0.33$, then we conclude that node $n$ fails longer term. 
This cutoff value is based on the report for Oak Ridge National Laboratory\cite{kappenman2010geomagnetic}, in which we can find that \textit{``the standard approach to spares has been to purchase an extra single phase transformer for a three phase bank''}.
Thus, we assume that if the failure rate is more than $33\%$ in a given region, the region suffers an extended power outage.

\subsection{Stage III}
We conclude how transformer failures affect population access to electricity. 
Our consideration here is based on the tessellation of the continents. 
It could also be performed in countries and alike, but we focus on continents in the analysis.
Unfortunately, the region served by each substation node is not publicly available, so we use a proxy approach -- a Voronoi diagram.
Each power grid node in the power distribution data serves as the centre for a Voronoi cell.
This way, we have a good approximation of real-world electric districts since each power substation serves its nearest surroundings.
Naturally, we limit the Voronoi regions to the extent of the geographic boundaries of the networks.
We use the $15$ arc-minute grid population density data to estimate the population in each cell \cite{popdata}.
Specifically, in each grid cell, the land area was multiplied by the population density to obtain the absolute population estimates. 
The total population in each Voronoi region was calculated as the total number of centres of $15$ arc-minute population grid cells within the Voronoi polygon represented by a power node. 

We assume the power outages to be long-term as the grid is often run at near-maximum capacity, and it is estimated that a region with insufficient transformers would be unable to serve electricity for
many months until those transformers were replaced, which is a slow and expensive process. This is a lower bound estimate, however, and could well take much longer \cite{kappenman2010geomagnetic}.

In addition to the number of people being cut off from power, we provide estimates on the total electric power loss in [GW]:
\begin{equation}
    P_{loss} = \frac{P_{cons}}{N} \sum_n C(n) [F(n) > 0.33].
\end{equation}
$P_{loss}$ is power lost, $P_{cons}$ is total power consumption (for continent/country, etc.) \cite{electrcountry}, $N$ is total population \cite{popcountry}, $C(n)$ is the population in a given Voronoi region $n$, $F(n)$ is from Eq.~\eqref{eq:failure}, and $[\dots]$ is the Iverson bracket notation.

\section{Results and Discussion}
\begin{figure*}[t]
    \centering
    \includegraphics[width=1\linewidth]{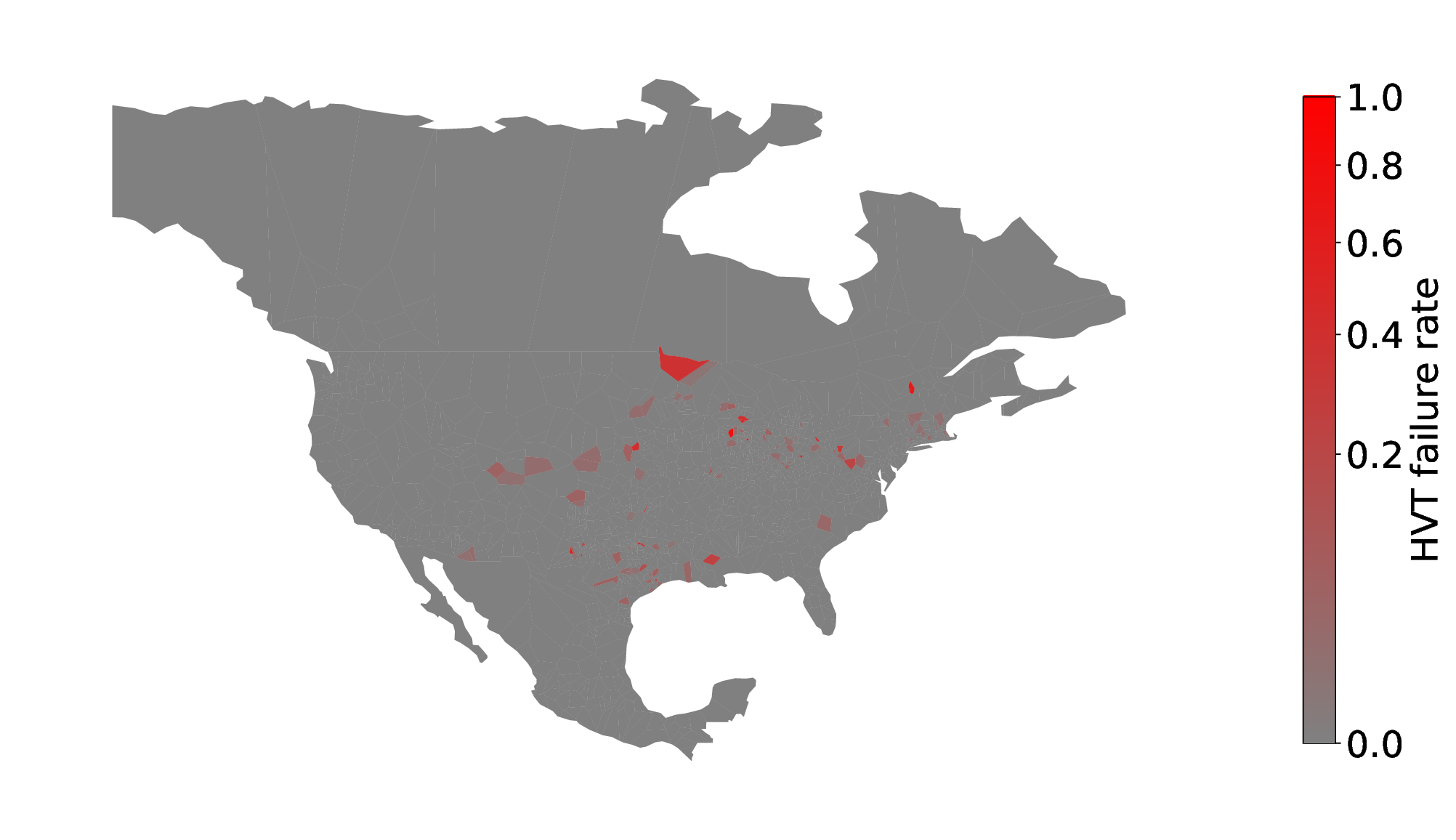}
    \caption{North American substations at risk of electricity loss after a $1$-in-$10000$ year storm. Colour indicates the fraction of HVTs we expect to fail within a given Voronoi tile.}
    \label{fig:navoronoi}
\end{figure*}

To recap, In Fig.~\ref{fig:flowchart}, we presented a simplified flowchart of our model.
The data-driven framework we propose runs in three stages estimating: 1) geoelectric fields for severe storms, 2) GICs and transformer overheating in high voltage power distribution networks, and 3) effects of transformer failure on the electric grid and population's access to electricity.
This three-stage process is an agglomeration of multiple smaller models we assembled across various scientific fields and modified or expanded where appropriate to suit our needs.

We utilise magnetotelluric (MT) \cite{gjerloev2009global} and EMTF \cite{kelbert2011iris,kelbert2019taking, kelbert2019first} site data to find geoelectric fields as described by Love et al.,~\cite{love2018geoelectric}.
Then, an adjustment process is applied to all fields to a reference location, with an intermediary step incorporating auroral boundary movement.
The reference adjustment is done by using the geomagnetic latitude data from Ngwira et al., \cite{ngwira2013extended}.
To this data, we fit a log-normal function that serves as our model for the recurrence of the geoelectric fields \cite{love2020some, love2018geoelectric}.
Finally, an impedance estimate is computed using Earth's ground conductivity model from Alekseev et al., \cite{alekseev2015compilation}.
With all those elements, we can compute $E$ field levels at any location on Earth for any storm rarity and duration.

This general model can be used for an arbitrary geographical region, including the whole globe, as long as power distribution network data are available.
The software implementation is written in Python and the source code is open and available for anyone to use and modify. It is also worth noting that our implementation is additionally accessible to a wide audience as it does not require an expensive computer cluster, and can be run on a reasonably modern machine in less than an hour on the scale of a continent or smaller.

In order to see how particular regions are affected by geomagnetic storms, we include the Voronoi diagram plots for both continents depicting a specific number of substations at risk in a granular manner; see Fig.~\ref{fig:euvoronoi} for Europe, and Fig.~\ref{fig:navoronoi} for North America in a $1$-in-$10000$ year scenario.
The colour indicates the ratio of HVTs within a given Voronoi cell that are expected to fail.
We acknowledge that Sweden has in fact hardened their grid to address the risks discussed here \cite{JRC86658}; however, for simplicity's sake, we include it in our results.

\begin{figure*}[t]
    \centering
    \includegraphics[width=1\linewidth]{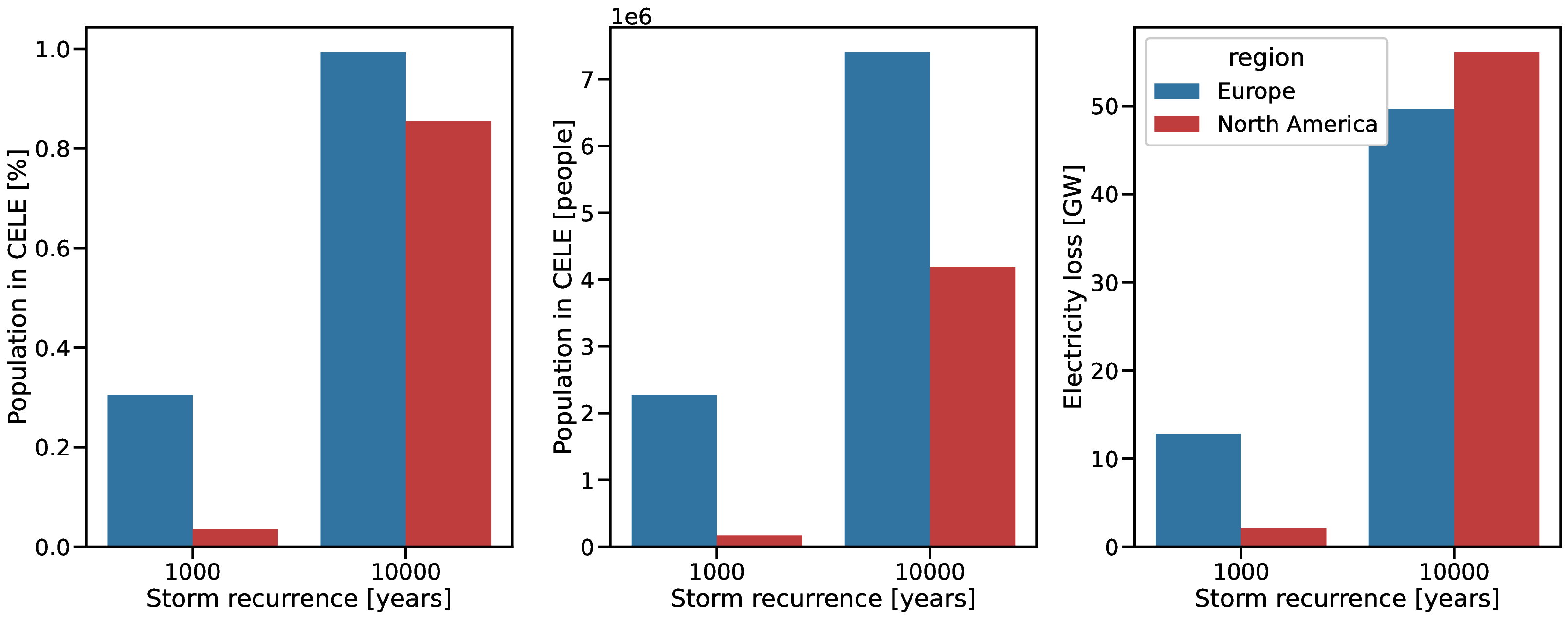}
    \caption{Population volume experiencing CEL as a function of storm rarity in $[\%]$ (left), the number of people (centre), and electricity loss in gigawatts (right).}
    \label{fig:cele}
\end{figure*}

We determine a region to be expected to suffer a power loss by considering the transformer failures exceeding available spare parts \cite{kappenman2010geomagnetic}.
Thus, we assume that if the failure rate is more than $33\%$ in a given region, the region suffers an extended power outage.
In Fig.~\ref{fig:cele}, we present the estimates of the population numbers experiencing a CEL as a result of a geomagnetic storm, and total electricity loss in gigawatts.
This plot illustrates how many people will lose power for months or perhaps even years when an extreme storm arrives.
This extreme event's strength is the highest expected to occur once per certain amount of years.
For instance, for an event of a strength that is expected to happen once in $10000$ years, in Europe, we expect $0.99\%$ (over $7.4$ million people) of the population to experience CEL, whereas in North America, it would be $0.86\%$ (over $4.1$ million people).
The total expected loss is $49.71$ [GW] and $56.12$ [GW], respectively.
We limited ourselves here to these particular metrics, but of course, this analysis can be expanded to include the expected loss of GDP, industry production, and so on.

Geomagnetically induced currents are a serious cause for concern, considering how reliant many of humanity's critical infrastructure systems are on electric power \cite{foster2008report}.
While plenty of papers tackle this issue in a localised and narrow manner, to our knowledge, no attempt at a global risk analysis has been conducted.
We tackle the problem of filling this knowledge gap and develop a general model that combines all adequate specific methods from thermal models of HVT tie bar hot spots through historical geoelectric field estimates to a global conductivity model.
It is our belief that the methodology presented in this paper can be applied on a global scale and inform the decisions of policymakers in regulating power distribution network resilience.

Our main analysis was limited to only two continents -- Europe and North America -- due to a lack of quality data in other regions; however, these two serve as very good proofs of concept.
The key results show that a severe geomagnetic storm can severely affect millions of people.
We believe that rather than approaching our results as final, they should be considered as a starting point for further analysis and a powerful tool and framework for future work, where more accurate grid connections, transformer designs and voltages, grounding resistances, and other specifics on a global scale can be used to obtain realistic results for a given region.
The unpredictable and rare nature of coronal mass ejections impacting Earth makes the risk assessment much more difficult since, at the time of writing this manuscript, it is virtually impossible to say what precisely the chances of a $1$-in-$10000$ year storm occurring within the next decade are.
While it seems prudent and most likely cost-effective to prepare the grid for severe storms instead of waiting for failures to happen, a global-scale complete power outage does not seem possible with the assumptions we have made in our model.

With that said, the data we have used to predict HVT overheating is limited. 
Our grid model relies on potentially incomplete or incorrect infrastructure data crowd-sourced by OpenStreetMaps, due to the lack of official public high voltage infrastructure data. 
However, this does not represent a failure of the model, as electrical utilities and government agencies can readily replace our grid map with their own private data for enhanced prediction accuracy.

One could potentially adapt our model to use a characteristic power line length or a synthetic network instead of real data, but this leads to the problem of losing regional precision in predicted transformer outages.
We would still have a global estimate of expected power loss, but the exact locations of predicted outages would be unreliable.

Similarly, the magnetotelluric data are still very limited, and while our method of normalising these data to a reference location and, from there, extrapolating worldwide has the advantage of universality, it does come with a drawback of potentially over-generalisation.
This can lead to underestimating storm recurrence in some specific regions.

Most importantly, we only consider transformer failure due to overheating; however, the safety switch cascade failure scenario is also a very important avenue of investigation that should be explored in the future, as well as other more complex interactions with the increasingly complex digital infrastructure automatically routing power through the electrical grid.

Overall, to our knowledge, the approach proposed in this paper is the first of its kind in assessing power distribution network resilience to GICs and can be broadly applied at municipality, country, continent or even a global level to inform governing bodies and private entities alike of the dangers associated with transformer overheating during geomagnetic storms.

\section*{Disclosure statement}
Authors report that there are no competing interests to declare.

\section*{Data availability statement}
All data used in this research are publicly available. The program source code and instructions on how to procure the data can be found in the github repository of the project: \href{https://github.com/allfed/GeomagneticModel}{https://github.com/allfed/GeomagneticModel}.

\nocite{*}

\bibliography{ms}

\end{document}


\maketitle

\newpage
\section{Methodology, stage one}
\begin{figure}[!htb]
    \centering
    \includegraphics[width=\textwidth]{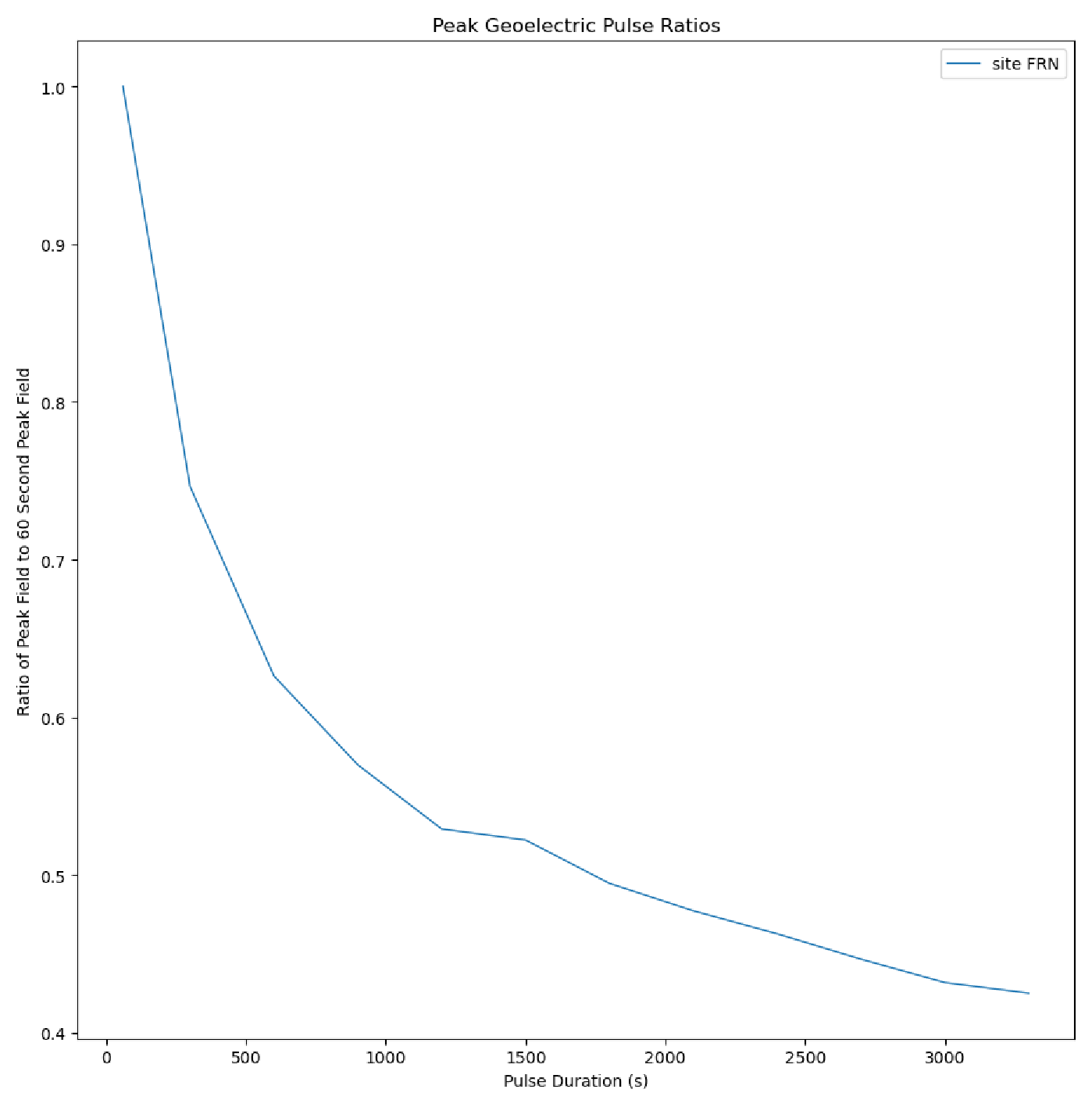}
    \caption{Peak $E$ field level as a function of the field duration, measured at the FRN MT site. Plot shows the field level as a ratio to the field strength at $60$ seconds.}
    \label{fig:duration}
\end{figure}
\begin{figure}[!htb]
    \centering
    \includegraphics[width=\textwidth]{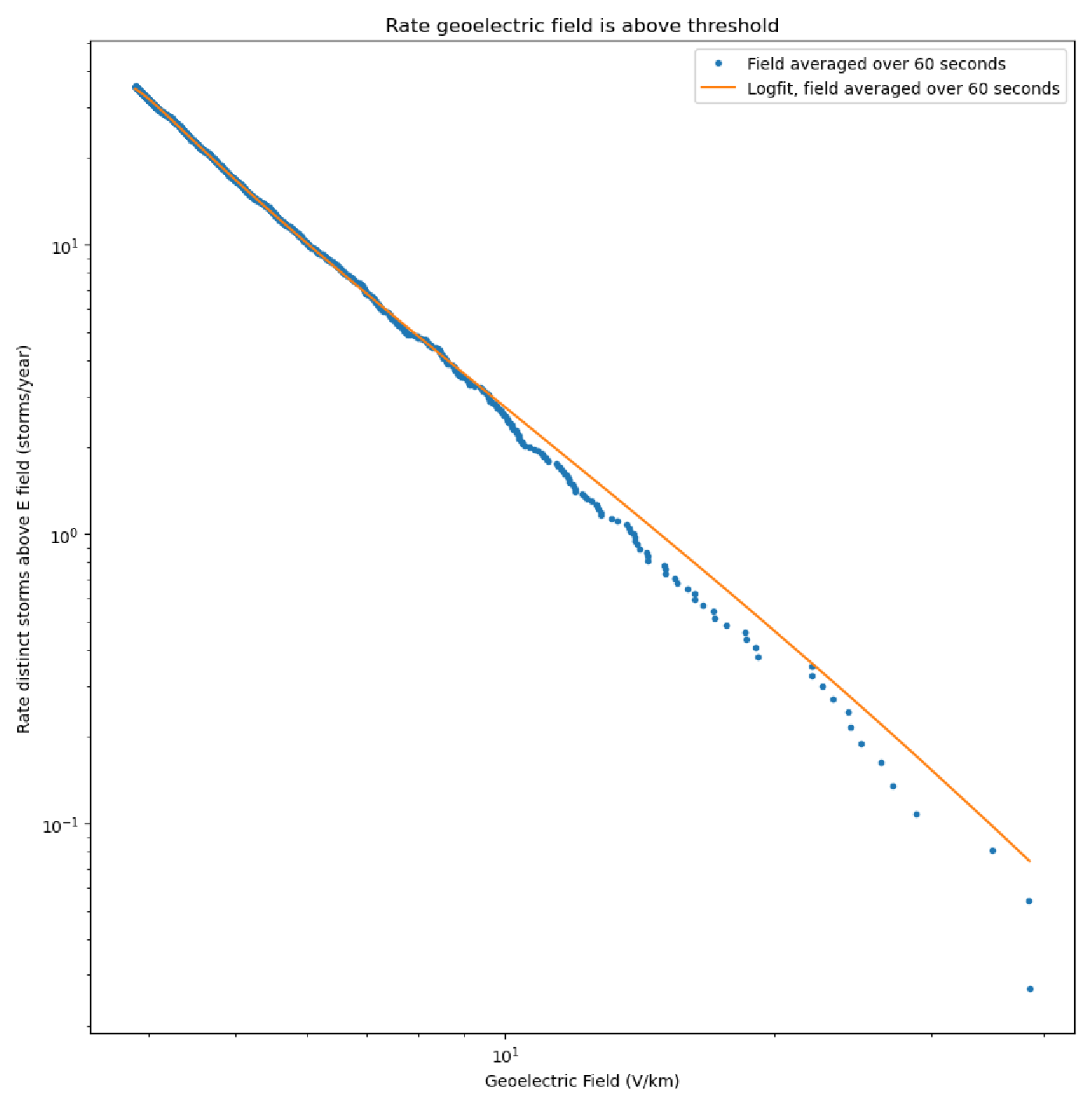}
    \caption{Normalised recurrence rates of geoelectric fields measured at the FRN MT site. This is the final result of our normalisation procedure in stage one of the method. We use the log-normal fit as per the recommendation in \cite{love2020some, love2018geoelectric}.}
    \label{fig:efieldfit}
\end{figure}

\newpage
\section{Methodology, stage two}
\begin{figure}
    \centering
    \includegraphics[width=\linewidth]{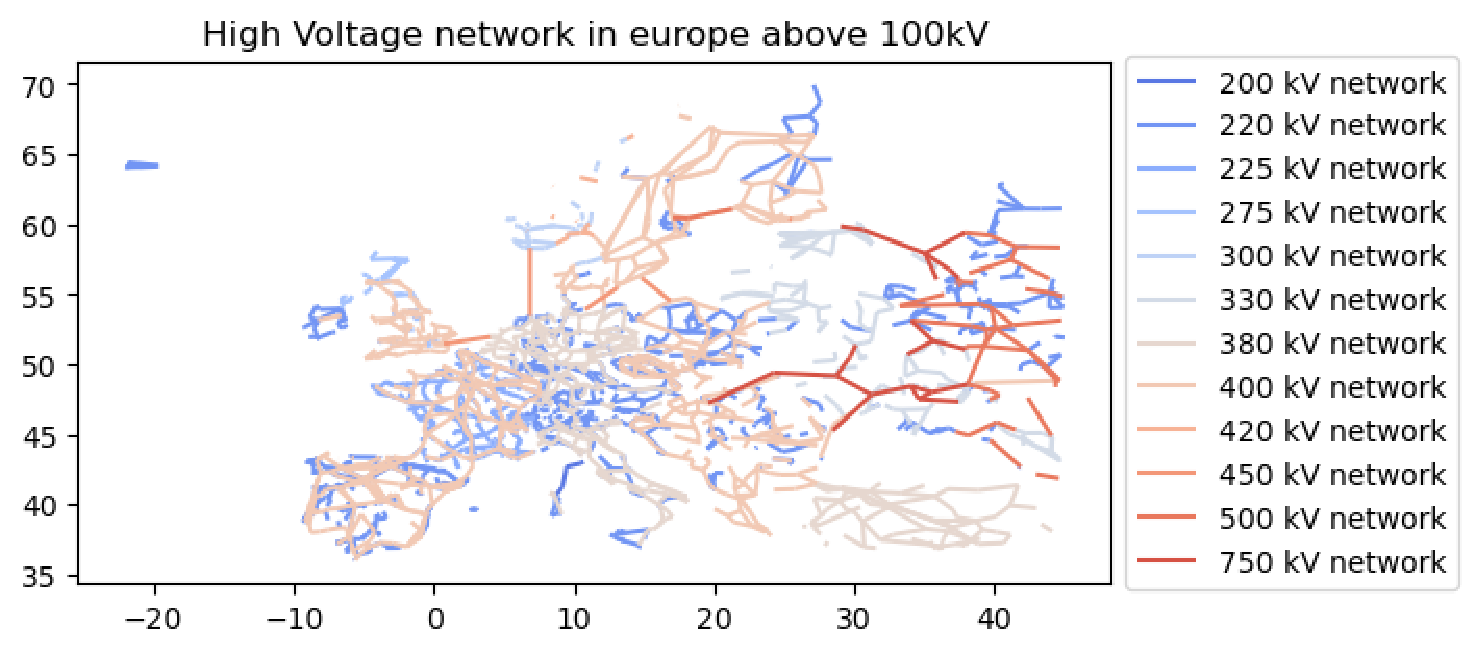}
    \includegraphics[width=\linewidth]{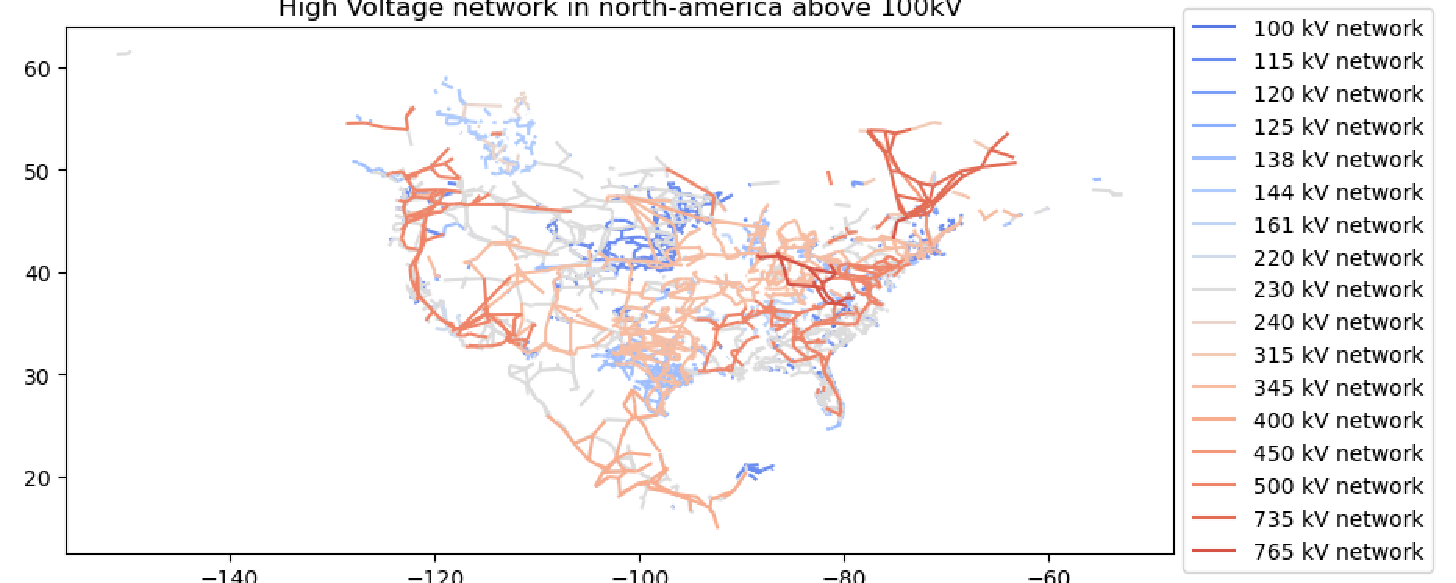}
    \caption{Power distribution networks for Europe (top) and North America (bottom) as acquired from OpenGridMap \url{https://github.com/OpenGridMap/transnet-models/tree/master/europe}. Only lines above $100[kV]$ are shown.}
    \label{fig:grids}
\end{figure}

\begin{table}[!htb]
    \centering
    \begin{tabular}{c|c|c|c|c}
    name    &voltage    & $V_c$ &$\phi$ & $f(V_c)$ \\
\hline
Tie1    &526    &500    &single & 1/7 \\
Tie2    &525    &500    &single & 1/7 \\
Tie3    &525    &500    &single & 1/7 \\
Tie4    &525    &500    &three  & 1/4 \\
Tie5&   525 &500    &three  & 1/4 \\
Tie6    &500    &500    &single & 1/7 \\
Tie7    &500    &500    &single & 1/7 \\
Tie8    &525    &500    &single & 1/7 \\
Tie9    &500    &500    &three  & 1/4 \\
Tie10   &525    &500    &three  & 1/4 \\
Tie11   &735    &765    &single & 1/3 \\
Tie12   &765    &765    &single & 1/3 \\
Tie13   &746    &765    &single & 1/3 \\
Tie14   &400    &345    &single & 1/7 \\
Tie15   &420    &345    &single & 1/7 \\
Tie16   &433    &500    &single & 1/7 \\
Tie17   &410    &345    &single & 1/7 \\
Tie18   &420    &345    &three  & 1/9 \\
Tie19   &400    &345    &three  & 1/9 \\
Tie20   &405    &345    &single & 1/7 \\
Tie21   &405    &345    &three  & 1/9 \\
Tie22   &400    &345    &three  & 1/9 \\
Tie23   &420    &345    &three  & 1/9 \\
Tie24   &335    &345    &single & 1/7 \\
Tie25   &345    &345    &single & 1/7 \\
Tie26   &275    &230    &three  & 1/7 \\
Tie27   &275    &230    &three  & 1/7 \\
Tie28   &330    &345    &three  & 1/9 \\
Tie29   &345    &345    &single & 1/7 \\
Tie30   &345    &345    &three  & 1/9 \\
Tie31   &345    &345    &three  & 1/9 \\
Tie32   &345    &345    &three  & 1/9 \\
Tie33   &230    &230    &single & 1/5 \\
Tie34   &231    &230    &single & 1/5 \\
Tie35   &230&   230 &single & 1/5 \\
Tie36   &230    &230    &three  & 1/7 \\
Tie37   &230&   230 &three  & 1/7 \\
Tie38 &230& 230 & three & 1/7 \\
Tie39 & 242 &230&single&1/5 \\
Tie40&240&230&single&1/5 \\
Tie41&225&230&three&1/7 \\
Tie42&230&230&three&1/7\\
    \end{tabular}
    \caption{Transformer categories. $V_c$ is the voltage class from the main text, $\phi$ is phase, $f(V_c)$ is the fraction for the transformer type. Values acquired from \cite{epri2019}.}
    \label{tab:hvtcategories}
\end{table}

\begin{table}[!htb]
    \centering
    \begin{tabular}{c|c|c}
         
Temp limit [$^\circ C$] & Age (years) & Transformer Population $f(A)$ \\
\hline
180 & 0-25 & 0.36 \\
160 & 25-40 & 0.25 \\
140 & $>40$ & 0.39
    \end{tabular}
    \caption{Transformer structural temperature limits per age category $T_{max}$ and their respective prevalence as fraction of transformer population $f(A)$. Values acquired from \cite{epri2017}.}
    \label{tab:hvtage}
\end{table}

\begin{table}[!htb]
    \centering
    \begin{tabular}{c|c|c|c|c}
         $V_c [kV]$ & Approx. Number in US Grid &  $f_{grid}(V_c)$ & $\phi$ & $f(\phi)$ \\
         \hline
230&1388&0.39&Single&0.25\\
230&1388&0.39&Three&0.75\\
345&1572&0.44&Single&0.15\\
345&1572&0.44&Three&0.85\\
500&587&0.16&Single&0.66\\
500&587&0.16&Three&0.34\\
765&58&0.01&Single&0.97 *\\
765&58&0.01&Three&0.03 *
    \end{tabular}
    \caption{Transformer population in US grid and fractional population by phase for each voltage class. Values marked with * were rounded to the nearest integer because no 765[kV] transformers were present in the other data sets. Values acquired from \cite{Kappenman2010MetaR322LP}.}
    \label{tab:hvtusgrid}
\end{table}

\bibliographystyle{ieeetr} 
\bibliography{refs}